# Brownian motion of molecules: a stochastic approach

Roumen Tsekov
Department of Physical Chemistry, University of Sofia, 1164 Sofia, Bulgaria

A stochastic Langevin equation is derived, describing the thermal motion of a molecule immersed in a rested fluid of identical molecules. The fluctuation-dissipation theorem is proved and a number of correlation characteristics of the molecular Brownian motion are obtained.

The main goal of this paper is to extend the region of validity of the classical theory of Brownian motion by demonstrating its applicability to problems related to the behavior of microscopic objects. The molecular diffusion process is theoretically studied, which in the frames of the kinetic theory of dilute gases can be described by the Boltzmann-Lorentz equation [1]. The solution of this problem by means of the methods of the classical theory of Brownian motion [2] requires derivation of a stochastic differential equation describing the motion of a molecule in a fluid and of the corresponding equation for the probability density evolution. Such a stochastic equation provides possibility to find numerous correlation characteristics of the process. A new feature following from the microscopic character of the present consideration is overcoming of the two main limitations of the classical theory of Brownian motion [2]; its phenomenological character and the application of the fluctuation-dissipation theorem (FDT).

## Stochastic theory of dilute gases

In the present paper our stochastic theory of dilute gases [3] is presented. The object of consideration is a fluid of mass points with mass $m$ being at thermodynamic equilibrium. The temperature of the system is denoted by $T$. At a given moment, accepted as initial, a particle identical to those of the fluid is immersed in the system. The basic aim is to describe the behavior of this target particle if its initial coordinate and velocity are known. Since the state of the fluid is given by average characteristics, the information about the target particle motion will possesses a statistical character. As was mentioned before, a tool for theoretical investigation of this self-diffusion process is the Boltzmann-Lorentz kinetic equation. However, it is a linear integro-differential equation governing the probability density evolution in the phase space only. The problem can also be attacked by the Fokker-Planck equation formalism [2] but this will provide just some general features of the target particle motion on a phenomenological level. Finally, the motion of the target particle can be described via a generalized Langevin equation (GLE) if proper expressions for the memory kernel are employed [2, 4].

In the present study the target particle motion is described on the basis of the theory of stochastic processes. A realization of the considered random process is a trajectory of the target particle in the thought experiment described above. Since all the realizations are independent, an operator for statistical average over the set of the realizations can be introduced via

$$<X(t)> \equiv \lim_{N \to \infty} \frac{1}{N} \sum_{n=1}^{N} X_n(t)$$

where $X_n$ is the realization of the random process $X$ observed in the *n*-th experiment. Using this operator every mean statistical characteristics of the process can be defined. For description of the behavior of a target particle via a stochastic dynamic equation it is necessary to know the forces acting on the particle. The basic concept of the present investigation is that the interaction can be modeled by impacts between the target and fluid particles. The notion of impact between two mass points is that for a small time interval denoted as collision duration the mass points receive finite changes in their linear momentums. Due to the usually small value of the collision duration, impact forces can be described sufficiently well without detailed knowledge about their complicate origin [5-7]. According to the laws of classical mechanics, the momentum and energy of an isolated system are constants. Hence, for an elastic impact between the target and a fluid particle the conservation laws read

$$m\dot{r}(t_s) + mv(t_s) = m\dot{r}(t_f) + mv(t_f) \qquad m\dot{r}^2(t_s) + mv^2(t_s) = m\dot{r}^2(t_f) + mv^2(t_f)$$

where $\dot{r}$ and $v$ are the velocities of the target and the fluid particles, respectively, $t_s$ and $t_f$ are the starting and final moment of interaction. Using the relation $\dot{r}(t_s) \neq \dot{r}(t_f)$, following from the existence of interaction [5], the system of energy and momentum conservation laws can be transformed into an equation for the change of the target particle velocity during the impact

$$\dot{r}(t_f) - \dot{r}(t_s) = \mathbb{E} \cdot [v(t_s) - \dot{r}(t_s)] \tag{1}$$

where $\mathbb{E}(t_s, t_f) = [\dot{r}(t_f) - \dot{r}(t_s)] \otimes [\dot{r}(t_f) - \dot{r}(t_s)] / [\dot{r}(t_f) - \dot{r}(t_s)]^2$.

The next step toward the solution of the problem is to express the full interaction force between the target particle and the fluid. Supposing relatively dilute fluids, the mean time between two consecutive impacts is longer than the collision duration. This means that practically the main contribution into the interaction is due to two-particle impacts. Hence, the resulting force $F$ can be considered as a superposition of two-particle collisions

$$F(t) = \int_0^t dt_s \int_t^\infty dt_f W(t_s, t_f) m[\dot{r}(t_f) - \dot{r}(t_s)]/(t_f - t_s)$$

where $W$ denotes the density of impacts starting at the moment $t_s$ and finishing at the moment $t_f$. In this expression the fact that only impacts starting before and finishing after the current time $t$ contribute to the resulting force, is accounted for. Substituting here eq. (1), the full force acquires the form

$$F(t) = \int_0^t ds\, \mathbb{G}(t,s) \cdot [v(s) - \dot{r}(s)] \tag{2}$$

where

$$\mathbb{G}(t,s) = \int_t^\infty d\tau\, W(s,\tau) m \mathbb{E}(s,\tau)/(\tau - s) \tag{3}$$

Generally, the target particle can be affected by an external potential $U$. However, due to the small duration of the collisions, the contribution of the external forces can be neglected in the collision description since they create negligible change into the particle momentum for this small time interval. Therefore, the two forces $F$ and $-\nabla U$ are additive and the Newton equation for the target particle reads

$$m\ddot{r} + \nabla U = F$$

Introducing here eq. (2), a generalized Langevin equation is obtained

$$m\ddot{r} + \int_0^t \mathbb{G}(t,s) \cdot \dot{r}(s) ds + \nabla U = \int_0^t \mathbb{G}(t,s) \cdot v(s) ds \tag{4}$$

In fact, eq. (4) looks like the ordinary GLE [2] but its memory function is a random quantity. It takes into account the random nature of the impact density, initial and final moments and direction of the resulting force. The further employment of eq. (4) requires an adequate expression for the impact density $W$ which determines the memory function via eq. (3). The simplest model, satisfying the presumed additivity of the interactions, is that of the hard sphere particles. This model implies zero duration of the impacts. In this extremely simple case the collision density is given by $W(s,\tau) = 4\sum \delta(s - t_i)\delta(\tau - t_i)$ [6], where $\{t_i\}$ is the set of all moments

of impacts between the target and fluid particles. The corresponding memory function is

$$\mathbb{G}(t,s) = 4\delta(t-s)m\mathbb{E}(t,t)\sum \delta(s-t_i)$$

As is seen, there is a Dirac delta function for the dependence of $\mathbb{G}$ on the difference $(t-s)$ just like it is in the case of a heavy Brownian particle with a white noise. Introducing the last expression into eq. (4) yields

$$m\ddot{r} + m\mathbb{B}\cdot\dot{r} + \nabla U = m\mathbb{B}\cdot v \tag{5}$$

where $\mathbb{B}(t) = 2\mathbb{E}(t,t)\sum \delta(t-t_i)$.

Equation (5) is the desired stochastic equation governing the dynamics of a molecule immersed in a dilute gas of identical particles. The interaction between the target and fluid particles is separated into two forces: a friction force being proportional to the velocity of the target particle and a fluctuation force $f = m\mathbb{B}\cdot v$, which is expressed via observable quantities in contrast to the macroscopic theory. An important difference between eq. (5) and the ordinary Langevin equation is the additional randomness of the friction force due to stochasticity of the specific friction tensor $\mathbb{B}$. As a whole the stochastic elements are three: the moments of impact $\{t_i\}$, the direction of the resulting force given by $\mathbb{E}$ and velocity $v$ of the fluid particle. Generally, one can suppose that these three random factors are statistically independent, assumption which in the language of the probability theory reads

$$<f_1(t_i)f_2(\mathbb{E})f_3(v)> = <f_1(t_i)><f_2(\mathbb{E})><f_3(v)> \tag{6}$$

where $f_1$, $f_2$ and $f_3$ are arbitrary functions. Due to the isotropy of the considered system there is no preferred direction of the collision force and the average friction tensor equals to

$$<\mathbb{B}(t)> = b(t)\mathbb{I} \tag{7}$$

where $\mathbb{I}$ is unit tensor and the specific friction coefficient is given by $b = 2\sum <\delta(t-t_i)>/3$ [6].

### Statistical properties of the fluctuation force

The obtained dynamic equation (5) provides possibility to find the statistical properties of the fluctuation force without using FDT. The main assumption is that the force $f$ is not affected by the state of the target particle. Therefore, its statistical properties can be obtained from the simplest case of a free target particle being into the fluid so long that it is fully relaxed.

If the potential force is omitted, eq. (5) reduces to

$$\ddot{r} + \mathbb{B} \cdot \dot{r} = \mathbb{B} \cdot v \qquad (8)$$

Using eqs. (6) and (7) the following two equations can be derived from eq. (8)

$$<\ddot{r}(t)\ddot{r}(s)> + b(t)<\dot{r}(t)\ddot{r}(s)> + b(s)<\ddot{r}(t)\dot{r}(s)> = b(t)b(s)[<v(t)v(s)> - <\dot{r}(t)\dot{r}(s)>] \qquad (9)$$

$$<\ddot{r}(t)\dot{r}(s)> = b(t)[<v(t)\dot{r}(s)> - <\dot{r}(t)\dot{r}(s)>] \qquad (10)$$

To determine the statistical properties of the fluctuation force it is necessary to employ some knowledge about the equilibrium state of the fluid. First, all variables have to be described via stationary random quantities. The combination of this condition with the state at rest of the whole system leads to zero value of the mean velocity of the fluid particles, $<\dot{r}> = <v> = 0$. Another consequence is time independence of the specific friction coefficient $b$. Finally, the velocity autocorrelation function of the target particle acquires the form $\mathbb{C}(t-s) = <\dot{r}(t)\dot{r}(s)>$ and, due to the equipartition theorem and lack of preferred direction, the velocity dispersion is $\mathbb{C}(0) = (kT/m)\mathbb{I}$.

The derivation of eq. (5) shows that $v$ is the velocity of the fluid particle which collides with the target particle at the moment $t$. It is clear, that in two different moments $v$ will be the velocity of two different particles. Hence, its autocorrelation function can be expressed as

$$<v(t)v(s)> = \theta(t-s)<v(t)\dot{r}(s)> + \theta(s-t)<\dot{r}(t)v(s)> \qquad (11)$$

where $\theta$ is the Heaviside function [$\theta(\tau > 0) = 1$; $\theta(\tau = 0) = 1/2$; $\theta(\tau < 0) = 0$] and the identity of the fluid and target particles is accounted for. Using the properties mentioned above and eqs. (9)-(11), the following equation describing the evolution of the velocity autocorrelation function of the target particle is obtained $\ddot{\mathbb{C}}(\tau) = b[\theta(-\tau) - \theta(\tau)]\dot{\mathbb{C}}(\tau)$. The solution of this equation under the obvious conditions $\mathbb{C}(0) = (kT/m)\mathbb{I}$ and $\mathbb{C}(\infty) = 0$ is the well-known in the theory of Brownian motion exponentially decaying function [2, 5, 6]

$$\mathbb{C}(\tau) = (kT/m)\mathbb{I}[\theta(\tau)\exp(-b\tau) + \theta(-\tau)\exp(b\tau)] = (kT/m)\mathbb{I}\exp(-b|\tau|)$$

Replacing it into eq. (10) and using eq. (11) one derives the velocity autocorrelation of the impact fluid particle

$$<v(t)v(s)> = 4(kT/m)\mathbb{I}\theta(t-s)\theta(s-t) = (kT/m)\mathbb{I}\delta_{ts} \qquad (12)$$

The autocorrelation function $<f(t)f(s)>=2kTmb\mathbb{I}\delta(t-s)=kT<\mathbb{G}(t,s)>$ of the fluctuating force $f=m\mathbb{B}\cdot v$ and some other statistical properties, e.g. $<f>=0$ and $<f(t>s)\dot{r}(s)>=0$, can be obtained from eq. (12). These relations are in agreement with the theory of GLE [2], which is a confirmation of the assumptions made. As is seen, the fluctuation force is delta correlated random quantity with zero mean value and divergent dispersion. The last shortcoming is due to the assumption that the duration of the collisions is zero, which leads to forces creating finite jumps in the particle momentum for infinitely small time intervals.

## Statistical properties of a target particle

According to eq. (8) the velocity of the target particle obeys the following expression

$$\dot{r}(t)=\exp[-\int_0^t \mathbb{B}(s)ds]\cdot \dot{r}(0)+\int_0^t \exp[-\int_s^t \mathbb{B}(\tau)d\tau]\cdot \mathbb{B}(s)\cdot v(s)ds \qquad (13)$$

A number of statistical characteristics of the Brownian behavior of the particle can be obtained by eq. (13) if the statistical properties of the fluctuation force are known. Averaging eq. (13) yields that the relaxation of the mean velocity of the target particle is given by the equation

$$<\dot{r}(t)>=\exp[-\int_0^t b(s)ds]\dot{r}(0)$$

where an equality proven in the Appendix is employed. It shows that the target particle loses its mean velocity exponentially with the number of undergone impacts and after sufficiently large time forgets the initial velocity value. This result is exactly the same as the behavior of a heavy Brownian particle in a rested fluid.

Another statistical characteristic of the target particle, the velocity autocorrelation function $\mathbb{C}(t,s)$ can also be derived from eqs. (12) and (13) and the Appendix. The result

$$\mathbb{C}(t,s)=(kT/m)\mathbb{I}\{\exp[-\int_0^t b(\tau)d\tau-\int_0^s b(\tau)d\tau]+\theta(t-s)[1-\exp(-2\int_0^s b(\tau)d\tau)]\exp[-\int_s^t b(\tau)d\tau]+$$

$$\theta(s-t)[1-\exp(-2\int_0^t b(\tau)d\tau)]\exp[-\int_t^s b(\tau)d\tau]\}$$

reduces to that obtained before for an equilibrium gas in the limit $t,s\to\infty$ at finite difference $(t-s)$. Finally, from eqs. (12) and (13) the statistical correlation between the velocities of the target and fluid particle can be obtained

$$<\dot{r}(t)v(s)> = (kT/m)\mathbb{I}[1+2\theta(s-t)]\theta(t-s)\exp[-\int_s^t b(\tau)d\tau]$$

The main difference between the correlations derived above and those obtained in the macroscopic theory of Brownian motion is the time dependence of the friction coefficient $b$. While a macro-particle undergoes large amount of simultaneous impacts, in the motion of a micro-particle there are time intervals when it does not interact with the surrounding media. This more specific effect can be taken into account via a time dependent friction coefficient.

## Conclusions

In the present paper a stochastic differential equation describing the motion of a molecule in a dilute fluid is derived. The statistical properties of the fluctuating force are obtained without using FDT. On the basis of this equation some correlation characteristics of the diffusion process are determined, where the discrete nature of the process is accounted for via a time dependent friction coefficient. Due to the formal identity of the stochastic equations governing the motion of a micro- and a macro-particles immersed in a fluid, the description of the microscopic process via the methods of the classical Brownian motion theory is possible.

## Appendix

The aim of the present part is to prove the equality

$$<\exp[-\int_s^t \mathbb{B}(\tau)d\tau]> = \exp[-\int_s^t b(\tau)d\tau]\mathbb{I} \tag{A1}$$

The proof is based on the stochastic equation (8) and the statistical properties of its random quantities, $<v>=0$ and $<\mathbb{B}>=b\mathbb{I}$. The formal solution of eq. (8) reads

$$\dot{r}(t) = \exp[-\int_0^t \mathbb{B}(\tau)d\tau]\cdot\dot{r}(0) + \int_0^t \exp[-\int_s^t \mathbb{B}(\tau)d\tau]\cdot\mathbb{B}(s)\cdot v(s)ds \tag{A2}$$

From this solution using the relation (6) and the statistical properties of $v$, one obtains the following expression for the average velocity of the target particle

$$<\dot{r}(t)> = <\exp[-\int_0^t \mathbb{B}(\tau)d\tau]>\cdot\dot{r}(0) \tag{A3}$$

On the other hand, this quantity can be derived by direct averaging of the stochastic equation (8) which leads to $<\ddot{r}>+b<\dot{r}>=0$. Hence,

$$<\dot{r}(t)>=\exp[-\int_0^t b(\tau)d\tau]\dot{r}(0) \tag{A4}$$

Combining expressions (A3) and (A4) one obtains the equality

$$<\exp[-\int_0^t \mathbb{B}(\tau)d\tau]>\cdot\dot{r}(0)=\exp[-\int_0^t b(\tau)d\tau]\dot{r}(0)$$

Since the initial velocity $\dot{r}(0)$ is an arbitrary quantity independent of the further evolution of the process, from the above equation follows the equality (A1).